\documentstyle[epsfig]{aa}

\def\rs{\rm s}
\def\rs1{\rm s^{-1}}

\def\rcm{\rm cm}
\def\rcm2{\rm cm^{-2}}

\def\cqr{\chi^2_\nu}
\def\c2nor{\chi^2}

\begin{document}


\title{Intrinsic spectra and energetics of BeppoSAX
Gamma--Ray Bursts with known redshifts}

\author{L.~Amati\inst{1}
\and F.~Frontera\inst{1,2}
\and M.~Tavani\inst{3}
\and J.J.M.~in 't Zand\inst{4}
\and A.~Antonelli\inst{5}
\and E.~Costa\inst{6}
\and M.~Feroci\inst{6}
\and C.~Guidorzi\inst{2}
\and J.~Heise\inst{4}
\and N. Masetti\inst{1}
\and E.~Montanari\inst{2}
\and L.~Nicastro\inst{7}
\and E.~Palazzi\inst{1} 
\and E.~Pian\inst{8}
\and L.~Piro\inst{6} 
\and P.~Soffitta\inst{6}
}

\offprints{L. Amati:amati@tesre.bo.cnr.it}

\institute{Istituto di Astrofisica Spaziale e Fisica cosmica - Sezione di Bologna,
CNR, Via Gobetti 101, I-40129 Bologna, Italy
\and
Dipartimento di Fisica, Universit\`a di Ferrara, Via Paradiso
 12, I-44100 Ferrara, Italy
\and
Istituto di Astrofisica Spaziale e Fisica cosmica - Sezione di Milano, CNR, Via Bassini 15, I-20133
Milano, Italy
\and
Space Research Organization Netherlands,
 Sorbonnelaan 2, 3584 CA Utrecht, The Netherlands
\and
Osservatorio Astronomico di Roma, Via Frascati 33,
  I-00040 Monteporzio Catone (RM), Italy
\and
Istituto di Astrofisica Spaziale e Fisica cosmica, CNR, Via Fosso del Cavaliere,
  I-00133 Roma, Italy
\and
Istituto di Astrofisica Spaziale e Fisica cosmica - Sezione di Palermo, CNR,
Via La Malfa 153, I-90146 Palermo, Italy
\and
Osservatorio Astronomico di Trieste
Via G.B. Tiepolo 11, I-34131, Trieste, Italy
}

\date{Received; Accepted }

\markboth{Intrinsic spectra and energetics of BeppoSAX
GRBs with known redshift}{}

\abstract{
 We present the main
 results of a study of spectral and energetics properties of
 twelve gamma-ray bursts (GRBs) with redshift estimates.
 All GRBs in our sample were detected by BeppoSAX in a broad energy range
(2--700 keV). From the redshift estimates and the
good-quality BeppoSAX time--integrated spectra we deduce the main
properties of GRBs in their cosmological rest frames.
 All spectra in our sample
are satisfactorily represented by the Band model with no
significant soft X--ray excesses or spectral absorptions. We
find  a positive correlation between the estimated
total (isotropic) energies in the 1--10000 keV energy range
($E_{rad}$) and redshifts $z$.
Interestingly, more luminous GRBs are characterized also  by
larger peak energies $E_p$s of their $EF(E)$ spectra.
Furthermore,  more distant GRBs appear to be systematically
harder in the X--ray band compared to GRBs with lower redshifts.
We discuss how
selection and data truncation effects could bias our results and
give possible explanations for the correlations that we found.
}


\authorrunning{Amati et al.}
\titlerunning{Intrinsic spectra of BeppoSAX
GRBs with known redshift}{}

\maketitle

\section{Introduction}

In this paper we report and discuss the main spectral properties of the X--
gamma--ray emission from Gamma-Ray Bursts 
(GRBs) with known redshift detected by the BeppoSAX satellite (\cite{boella97}).
We base our work on data obtained from the co-aligned wide field
detectors on board BeppoSAX: the Gamma Ray Burst Monitor (GRBM, 40--700 keV,
\cite{frontera97}) and the two Wide Field Cameras (WFC, 2--28
keV, \cite{jager97}). The combination of these detectors makes
possible not only localizations for GRBs occurring in the
WFC 20$^\circ$ x 20$^\circ$ (FWHM) field of view,
but also reliable estimates of their spectra from 2 to
$\sim$700 keV (e.g., Frontera et al. 2000\nocite{frontera2000}).

\begin{table*}
\caption{Observed physical parameters of GRBs included in our
sample}
\begin{tabular}{llllllllll}
\hline GRB & P$_{X}$$^{(a)}$& S$_{X}$$^{(b)}$&$\Delta \,T_X$$^{(c)}$
&P$_{\gamma}$$^{(a)}$ &S$_{\gamma}$$^{(b)}$ & $\Delta \,T_{\gamma}$$^{(c)}$ & 
 S$_{\gamma,min}$ & Redshift $z$
 &Ref. \\
           & (2--28 keV) &(2--28 keV)&(sec.) &(40--700 keV) &(40--700 keV)&(sec.) & $^{(d)}$ & $^{(e)}$ & $^{(f)}$ \\
\hline
970228 & 0.25$\pm$0.03  & 3.2$\pm$0.2 & 75 & 3.7$\pm$0.1  & 11$\pm$1 & 80  & 
 0.20  & 0.695  (HG) & (1) \\
970508 & 0.050$\pm$0.011  & 0.83$\pm$0.04 & 25 &  0.34$\pm$0.01   & 1.8$\pm$0.3 & 20   &  0.29  & 0.835  (OT + HG) & (2) \\
971214 & 0.060$\pm$0.009  & 0.32$\pm$0.05 & 40 &  0.68$\pm$0.07  & 8.8$\pm$0.9 & 35 & 0.70  & 3.42   (HG) & (3)\\
980326 & 0.15$\pm$0.03  & 0.55$\pm$0.08 & 9 &  0.245$\pm$0.015  &  0.75$\pm$0.15 & 9& 0.09  & 0.9$-$1.1   (PH) & (4)\\
980329 & 0.21$\pm$0.07  & 4.3$\pm$0.3 & 40 &  3.1$\pm$0.1   & 65$\pm$5 & 25      &  0.75  & 2.0$-$3.9  (PH) & (5)\\
980613 & 0.019$\pm$0.006  & 0.27$\pm$0.05 & 40 &  0.16$\pm$0.04 & 1.0$\pm$0.2 & 20 &  0.14  & 1.096  (HG) & (6)\\
990123 & 0.21$\pm$0.02  & 9.0$\pm$0.1 & 100 &  17.0$\pm$5.0  & 300$\pm$40 & 100  &  1.18  & 1.6  (OT) & (7)\\
990510 & 0.17$\pm$0.03  & 5.5$\pm$0.2 & 120 &  2.47$\pm$0.21  & 19$\pm$2 & 75     &  0.39  & 1.619  (OT) & (8)\\
990705 & 0.18$\pm$0.03  & 5.4$\pm$0.2 & 60 &  3.7$\pm$0.1   & 75$\pm$8 & 42 &
 1.05    & 0.843   (XP + HG) & (9) \\
990712 & 0.66$\pm$0.04  & 5.2$\pm$0.1 & 45 &  1.3$\pm$0.1   & 6.5$\pm$0.3 & 20   &  0.50  & 0.43   (OT + HG) & (8) \\
000214 & 0.075$\pm$0.010  & 1.7$\pm$0.1 & 80 &  4.0$\pm$0.2  & 14.2$\pm$0.4 & 10   &  0.21  & 0.37$-$0.47  (XA) & (10) \\
010222 & 0.47$\pm$0.04  & 21.0$\pm$0.3 & 280 &   8.6$\pm$0.2  & 92.5$\pm$2.8 & 130   &  0.55  & 1.473   (OT) & (11) \\
\hline
\label{t:tab1}
\end{tabular}
\begin{list}{}{}
\item[$^{(a)}$]Detected 1s peak fluxes in units of 10$^{-6}$ erg cm$^{-2}$
s$^{-1}$.
\item[$^{(b)}$] Detected fluences in units of 10$^{-6}$ erg cm$^{-2}$.
\item[$^{(c)}$] Evaluated from the background subtracted 1 s GRB light 
curve as the difference between the times of the last and first bins 
with a count $\approx 3\sigma$.
\item[$^{(d)}$]  Minimum detectable fluences (see text) in units of 10$^{-6}$ erg cm$^{-2}$. 
\item[$^{(e)}$] OT = redshift  determined from absorption lines in the
 Optical Transient spectrum; HG = redshift based on optical emission
lines of the  Host Galaxy spectrum; PH = redshift estimated from
photometric data; XP = redshift inferred from an absorption
feature in the X--ray spectrum of the prompt emission.
 XA = redshift inferred from an emission line in the spectrum of the
X--ray afterglow.
\item[$^{(f)}$]References for the redshift measurements: (1) Bloom et
al. 2000; (2) Bloom et al. 1998; (3) Kulkarni et al. 1998; (4) Bloom et
al. 1999; (5) Lamb et al. 1999 and references therein;
(6) Djorgovski et al. 2000; (7) Kulkarni et
al. 1999; (8) Vreeswijk et al. 2000; (9) Amati et al. 2000, Andersen et al. 2002;
(10) Antonelli et al. 2000;
(11) e.g. Stanek et al. 2001.
\end{list}
\end{table*}

Until now, 42 GRBs have been simultaneously detected by the
BeppoSAX WFC and GRBM. For twelve of these events, reliable
redshift estimates are available (see Table~\ref{t:tab1} and references
therein), allowing the investigation of systematic trends between
the GRBs spectral parameters and either GRB redshift or
total radiated energy. 
The extension of the spectral analysis to the X--ray
energy band allows a better determination of the continuum GRB spectrum, 
reducing the bias in the measurement of the spectral slope below the peak
energy $E_p$ of the $E F(E)$ spectrum.
Our analysis is based on  GRB time--integrated spectra. Compared
to the time-resolved spectra available for several of our GRBs,
the spectra considered in this paper  have a good statistical
quality, and are less affected by possible "absorption effects"
expected and observed at the very early times 
(\cite{frontera2000,amati2000}).

\section{The GRB sample and redshift estimates}

Table~\ref{t:tab1} summarizes the basic X-- and gamma--ray observed
properties of the GRBs included in our sample. For six events
(GRB~970228, GRB~970508, GRB~971214, GRB~980613, GRB~990705 and GRB~990712)
redshifts are available from optical emission lines in the
spectrum of their host galaxies.  For GRB~990123, GRB~990510, and
GRB~010222, lower limits are available  from the detection of
absorption lines in optical transient (OT) spectra. These lower
limits are reliable distance indicators: when redshift
measurements from both absorption and emission lines are
available, it is found that they are coincident.
For three more bursts (GRB~980326, GRB~980329 and GRB~000214) 
we adopt $z$
values based, respectively, on the time behavior of the optical afterglow, on optical/infrared
photometric
estimates and on the detection of an emission line in the X--ray afterglow 
spectrum. (see references  in Table~\ref{t:tab1}).
In column 8 of Table~\ref{t:tab1} we report for each GRB 
the 40--700 keV fluence $S_{\gamma,min}$ below which the event would not have been triggered 
and thus detected, assuming the same duration, light curve shape and spectrum 
of the detected GRB and as background level that measured during
the GRB observation.

As can be seen from Table~\ref{t:tab1}, the GRB emission properties 
reported (X-- 
gamma--ray fluence,  1s peak flux and GRB durations) vary over a broad range. 
The dynamic range of the peak flux and fluence is two orders of magnitude 
in $\gamma$-rays and one order of magnitude in X-rays. No significant 
correlation is found between redshift and GRBs peak flux, fluence or duration. 
The time profiles of GRBs in our sample are either simple or complex and the
range of GRB duration is wide, from 9 to 122 seconds. All the events 
belong to the "long GRB class", with $T_{90}$ durations larger than 2 s 
according to the standard GRB classification. 
The long duration of the GRBs localized by BeppoSAX may be due to the settings of the GRBM 
on--board trigger logic, which favors the detection of events with $\Delta$T $>$ few seconds, as reported by Guidorzi (2002)\nocite{Guidorzi02}. Possible selection effects 
which could affect our sample are discussed in more detail in section 5. 

%
%
\begin{figure*}
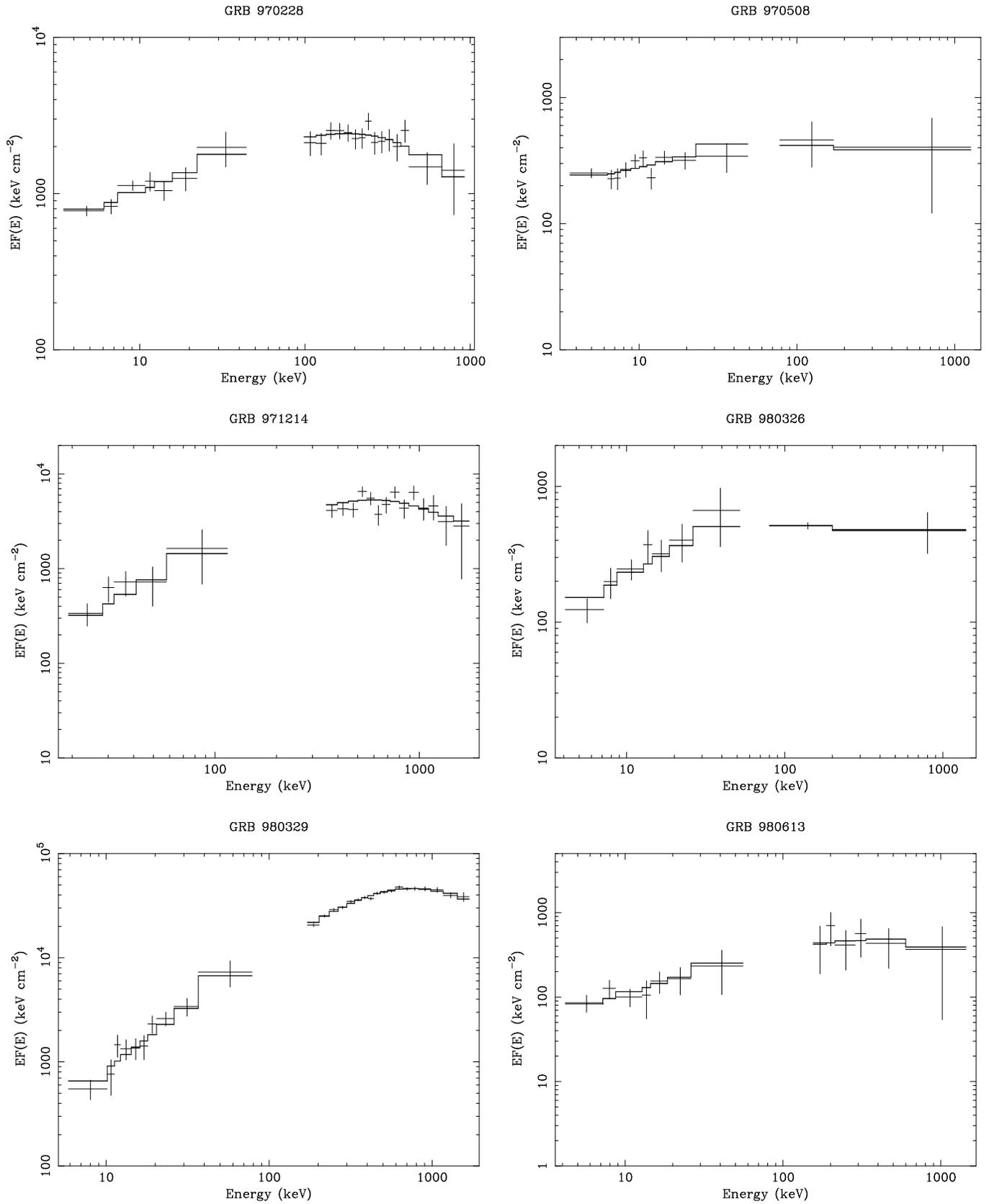

\centerline{\psfig{figure=970228_final.ps,width=7cm,angle=-90}\hspace{0.5cm}\psfig{figure=970508_final.ps,
width=7cm,angle=-90}}
\vspace{0.5cm}
\centerline{\psfig{figure=971214_final.ps,width=7cm,angle=-90}\hspace{0.5cm}\psfig{figure=980326_final.ps,
width=7cm,angle=-90}}
\vspace{0.5cm}
\centerline{\psfig{figure=980329_final.ps,width=7cm,angle=-90}\hspace{0.5cm}\psfig{figure=980613_final.ps,
width=7cm,angle=-90}}
\vspace{0.5cm}
\caption{Intrinsic time--integrated EF(E) spectra of the GRBs included in our sample.
WFC data: low--energy data--set; GRBM data: at higher energies.
Continuous line: best fit curve with the Band law (see Table~\ref{t:tab2}).}
\label{f:spectra}
\end{figure*}

\setcounter{figure}{0}
%
%
\begin{figure*}
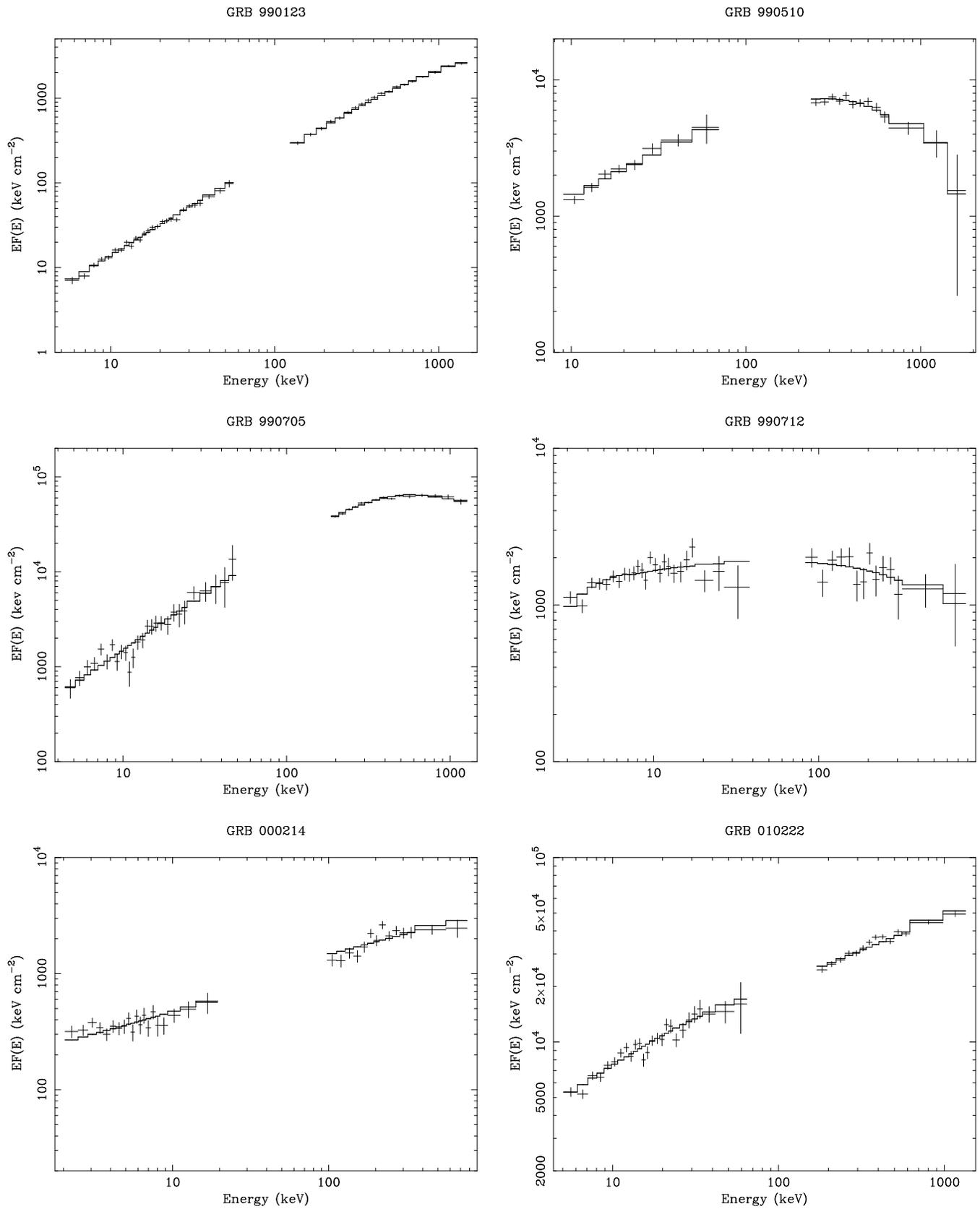

\centerline{\psfig{figure=990123_final.ps,width=7cm,angle=-90}\hspace{0.5cm}\psfig{figure=990510_final.ps,
width=7cm,angle=-90}}
\vspace{0.5cm}
\centerline{\psfig{figure=990705_final.ps,width=7cm,angle=-90}\hspace{0.5cm}\psfig{figure=990712_final.ps,
width=7cm,angle=-90}}
\vspace{0.5cm}
\centerline{\psfig{figure=000214_final.ps,width=7cm,angle=-90}\hspace{0.5cm}\psfig{figure=010222_final.ps,
width=7cm,angle=-90}}
\vspace{0.5cm}
\caption{(continued)}
\label{f:spectra}
\end{figure*}

\section{Data analysis method}

The spectral analysis was performed by adopting
standard GRBM and WFC data reduction techniques (see \cite{amati99a}
for the GRBM and \cite{jager97}) for the WFC. The
cross-calibration of the two instruments was performed using
simultaneous observations of the Crab source
(\cite{amati99b}). Given that the GRBM spectra have a time resolution
of 128 s, 
the effective exposure time to each event is estimated from 
the 1~s resolution light curves.
The spectra were fit assuming as input model the smoothly broken power--law 
proposed by
Band et al.\ (1993), whose parameters are the low--energy index
$\alpha$,  the break energy $E_0$ (in keV), the high energy index
$\beta$ and the normalization parameter (at 100 keV) $A$, and takes the
form:\\

$N(E) = A\cdot\left(\frac{E}{100keV}\right)^{\alpha}\cdot\exp{\left({-E/E_{0}}\right)}$
 , \\ 

$N(E) = A\cdot\left[\frac{({\alpha} - {\beta})\cdot E_{0}}{100keV}\right]^{\alpha - \beta}\cdot \exp{(\beta - \alpha)}\cdot \left(\frac{E}{100keV}\right)^{\beta}$ \\

for $E\hbox{  }\le\hbox{  } ({\alpha} - {\beta})\cdot E_{0}$
and
 for $E\hbox{  }\ge\hbox{  } ({\alpha} - {\beta})\cdot E_{0}$ respectively. 

In carrying out the
spectral fitting we took into account the Galactic 
absorption along the burst directions using the photo--electric cross--sections
by Morrison and Mc Cammon (1983) and the column density maps by
Dickey and Lockman (1990).

With respect to previous works (e.g. Jimenez et al. 2001, Bloom et al. 2001)
\nocite{Jimenez2001,Bloom2001}) we adopted a different
technique for the estimate of the intrinsic GRB properties.
Instead of fitting the observed spectra, estimating the GRB fluences and 
then applying a K--correction 
(method suggested by Bloom et al. 2001), we first blue-shifted the
GRB spectra to the their cosmological rest frames \footnote{If the
spectral analysis is performed using the XSPEC package (\cite{Arnaud96}),
as we did, the correction of the spectra for cosmological
redshift is obtained by multiplying by $(1 + z)$ the ENERG\_LO and ENERG\_HI columns
of the MATRIX extension and the E\_MIN and E\_MAX columns of the EBOUNDS extension
of the response matrix FITS files.
This task can be performed e.g. by using the $fv$ program belonging to
the FTOOLS data reduction and FITS file manipulation package (\cite{Blackburn95}).}
and thus we derive their
intrinsic shape. 
The total radiated energy
of a GRB in a fixed energy range is then simply computed by integrating the best--fit model in that
range (we adopt 1--$10^4$ keV) and scaling for source luminosity distance. 
This latter quantity is derived
assuming a flat Friedman-Robertson--Walker cosmological model with H$_0$ = 65 km s$^{-1}$ Mpc$^{-1}$,
$\Omega$$_m$ = 0.3, $\Omega$$_\Lambda$ = 0.7 (e.g. \cite{Carroll92}). For the sake of clarity, if 
N(E,$\alpha$,E$_0$,$\beta$,A) is the best fit Band model to the time--integrated and redshift--corrected spectrum of a GRB, D$_L$ is its luminosity distance and $z$ its redshift, the (1--$10^4$ keV)  total 
radiated energy $E_{rad}$ is given by:

$$E_{rad} = \frac{\int_{1}^{10000}{E N(E,\alpha,E_0,\beta,A) dE} \times 4 \pi D^2_L}{(1 + z)^2} $$

The $(1 + z)$$^2$ factor comes out from the fact that the luminosity distance is
defined in a way to account for the cosmological time dilation and spectral redshift when converting the observed
source flux to the source luminosity (e.g. \cite{Coles95}), and both effects depend on $(1 + z)$. Because we are dealing with fluences (i.e. the average flux multiplied to the GRB duration) and we have already corrected for spectral redshift before performing the spectral fitting, we have to divide by $(1 + z)$ twice.

Finally, the correlation
coefficients and the associated errors reported and discussed
in the next Sections were computed by 
properly weighing for data
uncertainties (e.g. \cite{Bevington69}) and verified with numerical simulations.

\begin{table*}[ht!]
\caption{Results of the Band model spectral
 fits of the 2$-$700 keV spectra transformed in the cosmological
 GRB rest frames~($^{\star}$) and derived total radiated energies assuming isotropic
emission.
}
\begin{tabular}{llccccccl}
\hline GRB & rest frame &  $\alpha$ & $\beta$ & 
$E_0$  & $E_p$ $^{(a)}$ & $A$ $^{(b)}$ & $\c2nor$/d.o.f. &  $E_{rad}$ $^{(c)}$ \\
    &  band (keV) &    &    & (keV)    & (keV)    \\
\hline
970228 & 3.4$-$1186  &  $-$1.54$\pm$0.08 & $-$2.5$\pm$0.4 & 424$\pm$33 & 195$\pm$64
& 0.331$\pm$0.013 & 19.1/18 & 1.86$\pm$0.14  \\
970508 & 3.7$-$1284 &  $-$1.71$\pm$0.10 & -2.2$\pm$0.25 & 502$\pm$150 & 
145$\pm$43 & 0.074$\pm$0.015 & 7.7/8  & 0.71$\pm$0.15 \\
971214 & 8.8$-$3094  &  $-$0.76$\pm$0.17 & $-$2.7$\pm$1.1  & 552$\pm$73 & 685$\pm$133
& 0.268$\pm$0.030 & 20.7/17 & 24.5$\pm$2.8  \\
980326 & 4$-$1400  &  $-$1.23$\pm$0.21 & $-$2.48$\pm$0.31 &  92$\pm$20 & 71$\pm$36
& 0.160$\pm$0.032 & 1.2/5 & 0.56$\pm$0.11  \\
980329 & 7.9$-$2765  & $-$0.64$\pm$0.14 & $-$2.2$\pm$0.8 & 687$\pm$80 & 935$\pm$150
& 1.36$\pm$0.32 & 26.5/24 & 210.7$\pm$20.3  \\
980613 & 4.2$-$1467 &  $-$1.43$\pm$0.24 & $-$2.7$\pm$0.6 & 342$\pm$170 & 194$\pm$89
& 0.072$\pm$0.012  & 2.9/11 & 0.68$\pm$0.11 \\
990123 & 5.2$-$1820  &  $-$0.89$\pm$0.08 & $-$2.45$\pm$0.97 & 1828$\pm$84 & 2030$\pm$161
& 2.23$\pm$0.22 & 21.5/41  & 278.3$\pm$31.5  \\
990510 & 5.2$-$1834  &  $-$1.23$\pm$0.05 & $-$2.7$\pm$0.4 & 549$\pm$26 & 423$\pm$42
& 0.897$\pm$0.091 & 19.7/17  & 20.6$\pm$2.9  \\
990705 & 1.71$-$1290  &  $-$1.05$\pm$0.21 & $-$2.2$\pm$0.1 & 366$\pm$13 & 348$\pm$28   &  1.61$\pm$0.18
 & 46.5/36  & 21.2$\pm$2.7   \\
990712 & 2.9$-$1001  &  $-$1.88$\pm$0.07 & $-$2.48$\pm$0.56 & 779$\pm$125 & 93$\pm$15 & 0.223$\pm$0.013
& 36.1/34 & 0.78$\pm$0.15   \\
000214 & 2.8$-$994  &  $-$1.62$\pm$0.13 & [-2.1]  & $>$308 & $>$117 & 0.185$\pm$0.006
& 57.1/37 & 0.93$\pm$0.03   \\
010222 & 4.9$-$1731 &  $-$1.35$\pm$0.19 & $-$1.64$\pm$0.02 & 146$\pm$41 & 
$>$886 $^{(d)}$ & 4.02$\pm$0.25
& 51.6/41 & 154.2$\pm$17.0  \\
\hline
\end{tabular}
\begin{list}{}{}
\item[$^{(\star)}$] 
For GRB~980326, GRB~980329 and GRB~000214 we assumed the central
value of the redshift intervals reported in Table~\ref{t:tab1}.
\item[$^{(a)}$]
$E_p = E_0 \times (2 + \alpha)$. 
\item[$^{(b)}$]
In photons cm$^{-2}$ keV$^{-1}$.
\item[$^{(c)}$]
Expressed in units of 10$^{52}$ erg.
\item[$^{(d)}$]
Lower limit computed by assuming $\beta$ = -2.1 (see also text).

\end{list}
\label{t:tab2}
\end{table*}

\section{Results}

Figure~\ref{f:spectra} shows the intrinsic time--integrated EF(E) spectra of the
GRBs included in our sample.
All but one (GRB000214) WFC plus GRBM redshift--corrected spectra
are well described by the Band model. 
For GRB~000214, the spectrum is well fit by a simple power--law up to the high
energy threshold of the GRBM; fitting this spectrum with the Band model
only a
3$\sigma$ lower limit on $E_p$ can be derived by fixing $\beta$
at $-$2.1. 
The
results of the fits   
are reported in Table~\ref{t:tab2}, where the
quoted uncertainties are 1$\sigma$ errors. 
For GRB~010222, $\beta$ is higher than -2, thus we have estimated a
lower--limit for the peak energy by assuming $\beta = -2.1$ . 
Table~\ref{t:tab2} also shows the total intrinsic released energies in
the 1--10$^4$~keV range.
For those events (namely, GRB 000214 and GRB010222) for which $E_p$ could not be determined, we derived the minimum and maximum values of $E_{rad}$  by 
assuming $E_p$ equal to its lower limit and 
to 10000 keV, respectively (while keeping fixed $\alpha$, $\beta$ and the 
normalization A). 

%
%
\begin{figure}
\hspace{-0.73cm}\psfig{figure=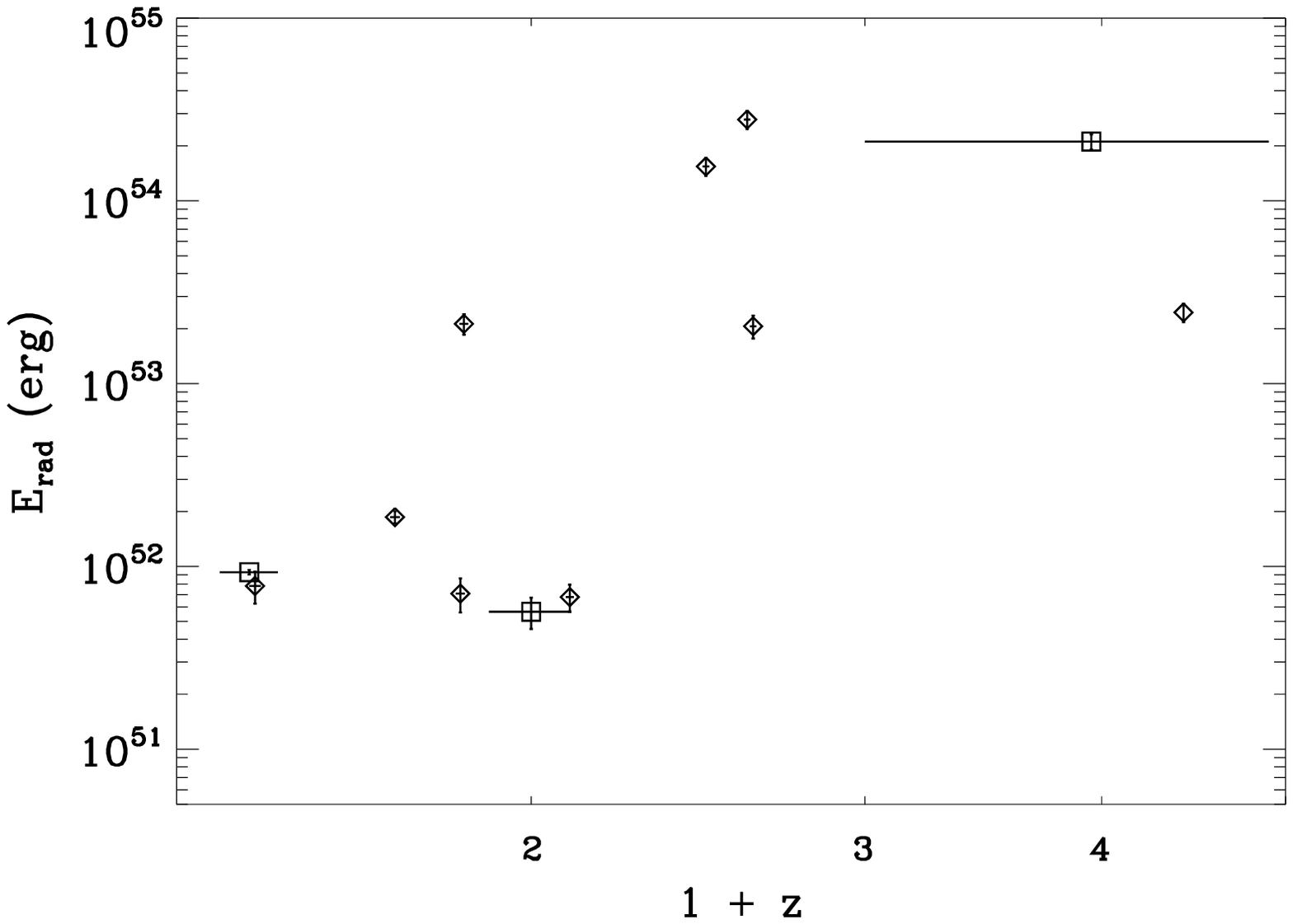,width=9.5cm,angle=0}
\caption{Dependence on redshift of the isotropic total
radiated energy.
Diamonds indicate the 9 GRBs with firm redshift
estimates, while squares refer to GRB~980326, GRB~980329 and GRB~000214.
}
\label{f:iso}
\end{figure}

%
%
\begin{figure*}[ht!]
\centerline{\psfig{figure=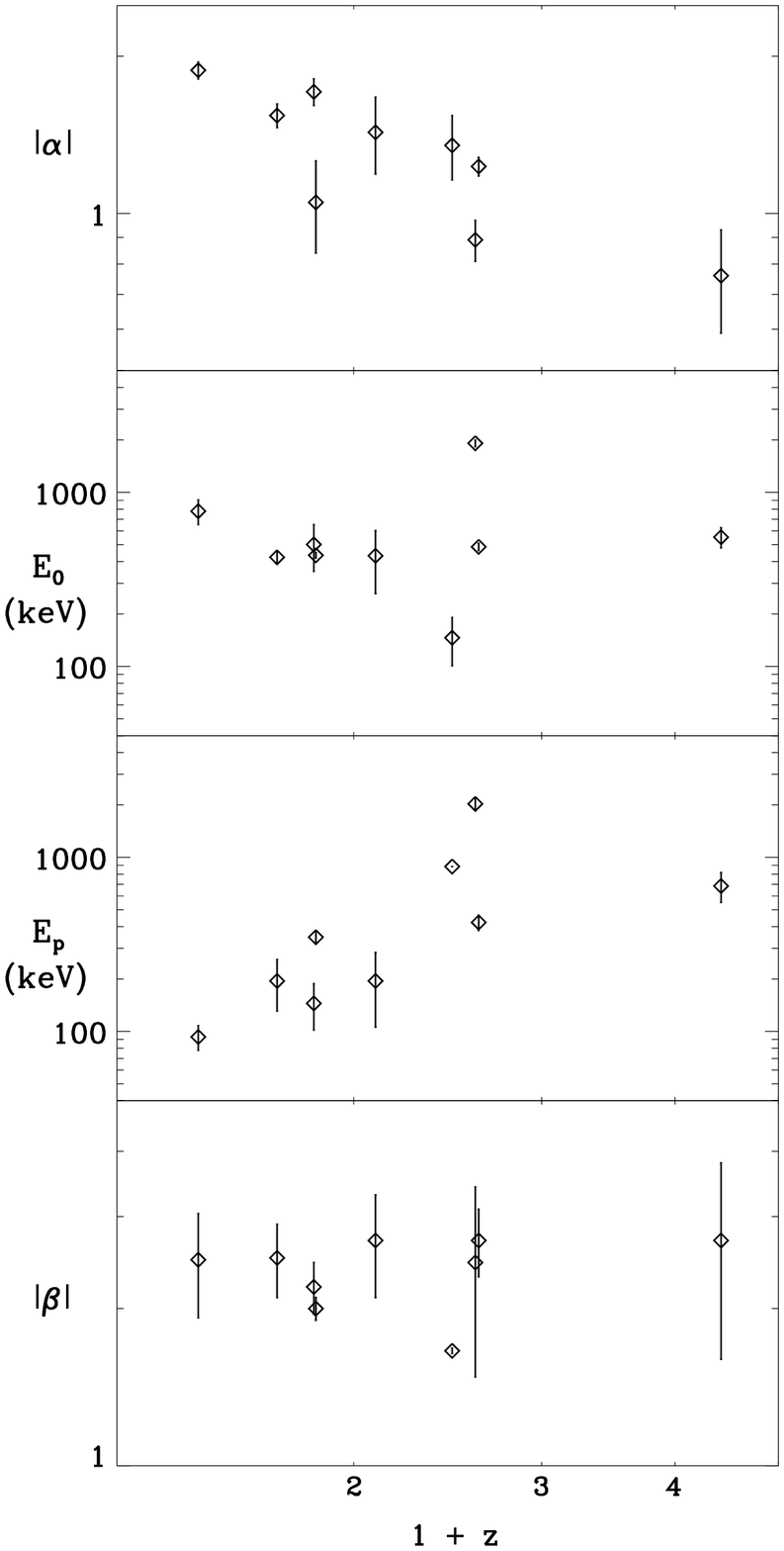,width=9.4cm,angle=0}
\hspace{-2.53cm}\psfig{figure=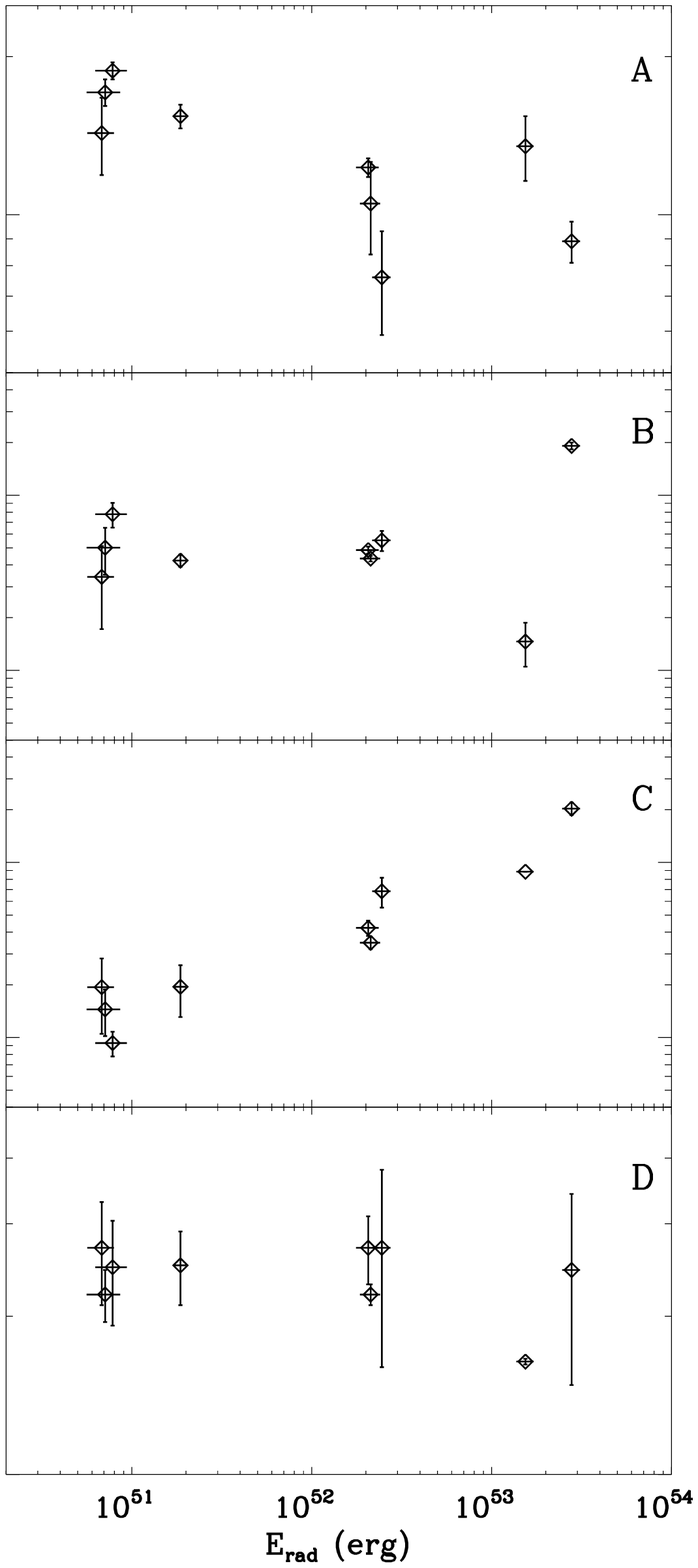,width=9.4cm,angle=0}}
\vspace{1.0cm} 
\caption{Dependence on redshift (left panels) and
isotropic total radiated energy (right
panels) of 
the X--ray photon
index (A), peak energy (B) and gamma--ray photon index (C) for the
nine GRBs with firm redshift determinations.}
\label{f:summary}
\end{figure*}

%
%
\begin{figure*}[th!]
\centerline{\psfig{figure=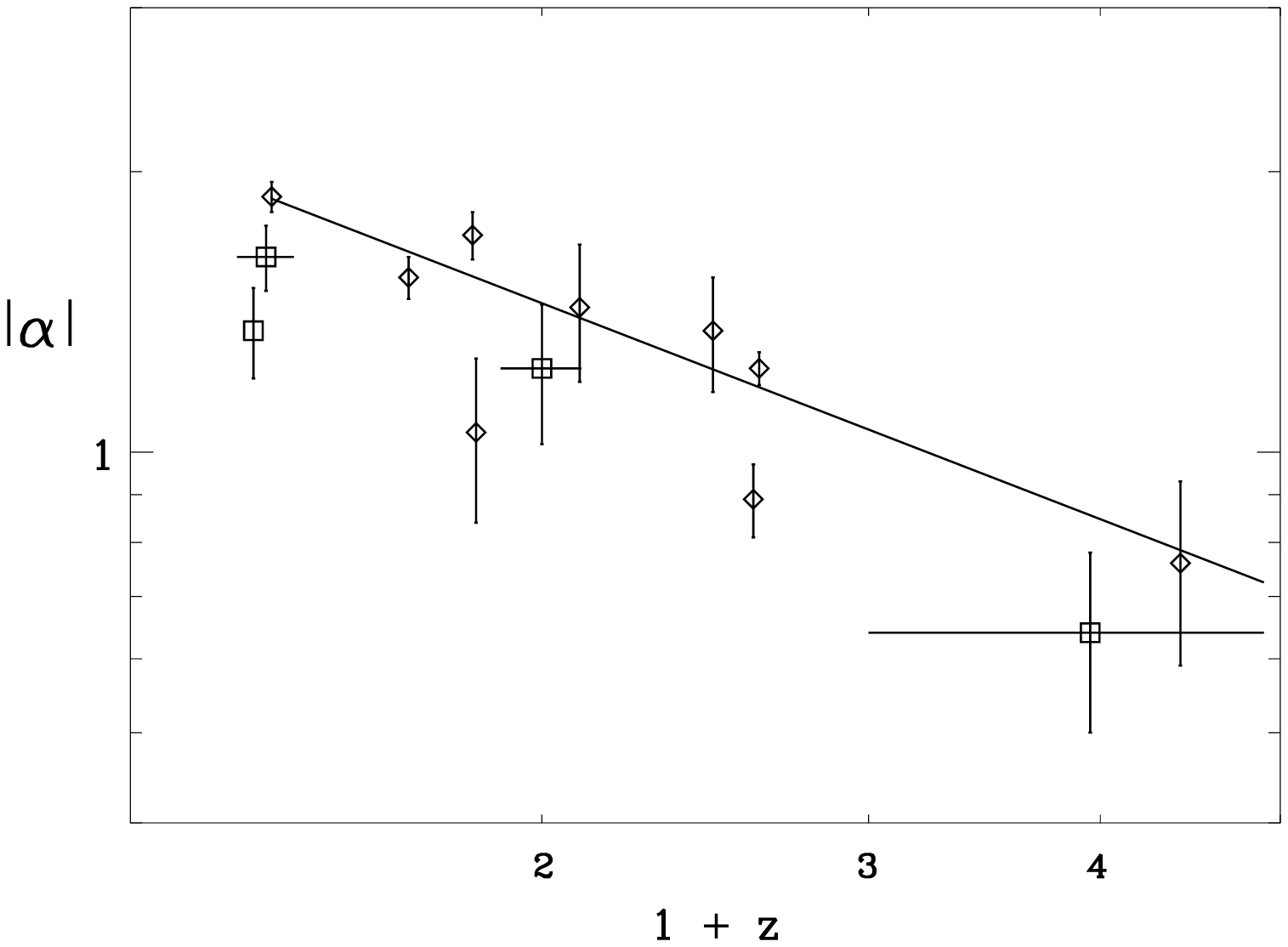,width=10cm,angle=0}
\hspace{-0.8cm}
\psfig{figure=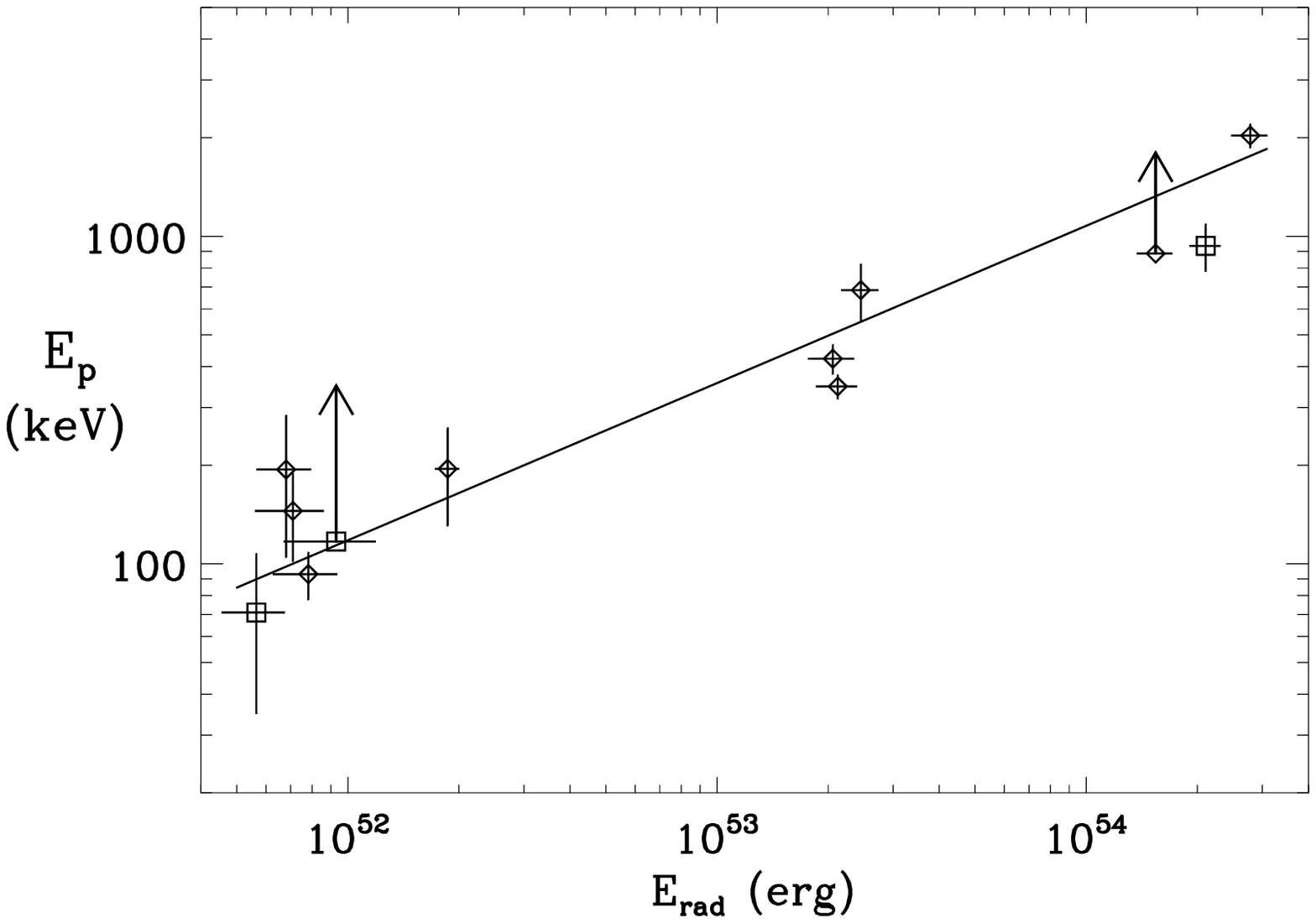,width=10cm,angle=0}}
\vspace{0.2cm}
\caption{Dependence on redshift of the X--ray
photon index (left) and dependence on isotropic total radiated energy of $E_p$ (right) for all the events included in the sample.
The symbols are the same used for Fig.~1. The right panel also includes the 
lower
limits on $E_p$ for GRB 000214 and GRB 010222.
 In each panel we plotted over the data the power--law
which best fits the values of the GRBs with
firm redshift estimate (diamonds).}
\label{f:detail}
\end{figure*}

In Fig.~\ref{f:iso} we show the total radiated energy $E_{rad}$ 
as a function of the redshift. 
Two important features can be inferred from these data:
a) a general trend of $E_{rad}$ to increase with $z$ 
(see also Table~\ref{t:tab3} for the correlation coefficient), and b)
the $E_{rad}$ values  of most of the events in our sample
are well above the sensitivity limit. 
We also find that the intrinsic X-- and $\gamma$--ray 
durations, computed by correcting the observed durations (see Table~\ref{t:tab1})  
for the $(1 + z)$ factor due to the cosmological time dilation, do not show  
significant correlations with redshift.

We performed a correlation study of the spectral parameters as a function of
redshift and total radiated energy for the 9 GRBs in our sample
with reliable redshift estimates.  
As clearly shown in Fig.~\ref{f:summary}, no correlation with redshift or
$E_{rad}$ is found for the high energy photon 
index $\beta$ and the break energy $E_0$ . 
However, we see an evidence of correlation with the redshift of the
low energy photon index $\alpha$ and of the peak energy $E_p$, and a clear
correlation between $E_p$ and
$E_{rad}$. 
The correlation between $\alpha$ and $E_{rad}$ appears to be weaker.
In Table~\ref{t:tab3} we report the correlation coefficients $r_{log}$ between
the logarithms of the quantities of interest. 
The correlation coefficients values between the direct quantities (not
reported) are generally lower, indicating that the correlations found are
better described by a power--law than a linear law.
The most significant correlations are (see Table~\ref{t:tab3})
between $log|\alpha|$ and $log(1 + z)$  and between
$logE_p$ and $logE_{rad}$ .  
Both correlations remain strong ($r_{log} = -0.842\pm 0.108$ and
$-0.902\pm0.078$,
respectively) even when the most distant event (GRB~971214), for the former, 
and the more energetic event (GRB~990123), for the latter, is excluded
from the data set. 
Including the three events with less firm redshift determinations (GRB~980326, 
GRB~980329 and GRB~000214), assuming for each of them the redshift given
by the centroid of the redshift interval reported in Table~\ref{t:tab1},
the $log|\alpha|$ vs. $log(1 + z)$ 
correlation is confirmed (r$_{log}$ = $-$0.859$\pm$0.078 for 12 events)
and the $logE_p$ vs. $logE_{rad}$ correlation is even more significant 
(r$_{log}$ = $-$0.941$\pm$0.037 for 10 events). The chance probability 
associated to these two correlations are 
0.1\% and 0.009\% respectively. 
Assuming a power-law for the relationship between  $|\alpha|$ and $(1 + z)$
and between $E_p$ and $E_{rad}$  we found the following best--fit slopes
(1 $\sigma$ uncertainties): $-0.78\pm 0.18$ ($\cqr$ = 0.54) and  
$0.52 \pm 0.06$ ($\cqr$ = 0.91), respectively.
The bets fit results are shown in Fig.~\ref{f:detail}, where 
we have also included the values corresponding to
GRB~980326, GRB~980329 and  GRB~000214 and
the lower limits on $E_p$ from GRB~000214 and GRB~010222.

\begin{table*}[ht!]
\caption{Correlation analysis results for the 9 GRBs with firm redshift estimates. The reported values and uncertainties have been computed following the method described in Section 3.  
The chance probabilities associated to the central values of the coefficients
are given in parenthesis for the more significant correlations. 
}
\begin{center}
\begin{tabular}{llll}
\hline
  &  ~~~~~log(1 + z) & ~~~~~logE$_{rad}$ & ~~~~~log$|$$\alpha$$|$    \\
\hline
\\
logE$_{rad}$      & ~~0.628$\pm$0.247 (7.0\%)  &                     &        \\
\\
log$|$$\alpha$$|$ & $-$0.859$\pm$0.091 (0.16\%) & $-$0.756$\pm$0.151 (1.6\%)       &     \\
\\
logE$_p$          & ~~0.636$\pm$0.184 (9.2\%)  & ~~0.949$\pm$0.036 (0.005\%)    & $-$0.821$\pm$0.075 (0.44\%)     \\
\hline
\end{tabular}
\end{center}
\label{t:tab3}
\end{table*}

\section{Discussion}

We derived the intrinsic broad-band (from few keV to few MeV, depending on source redshift) spectral properties and the total radiated energy in a fixed energy range (1--10000 keV)
of twelve BeppoSAX GRBs with known redshift. 
Thanks to the extension to X--rays of the spectral analysis the determination
of the low--energy spectral index of the Band function is less affected by
the spectral curvature when the peak energy is around 100--200 keV.
Determinations of the low--energy photon index  from the BATSE data
(e.g. \cite{Jimenez2001}) could be affected by these curvature effects.

First of all we find a trend of the isotropically released energy 
$E_{rad}$ to increase with $1+z$ (see Fig.~\ref{f:iso}).
This trend
is intrinsic and not due to sensitivity limitations.  In addition we find a
statistically significant correlation between the peak energy $E_p$ of the
$EF(E)$ spectrum and the energy $E_{rad}$. Also a correlation between the
X--ray photon index $\alpha$ and redshift is apparent. However we do not
find any evidence of a correlation between peak flux, fluence or duration and
redshift (see Table~\ref{t:tab1}).

Three possible explanations of these features can be considered:

(1) given the still small number of GRBs with known redshift, our sample 
is biased and thus is not representative of the overall
GRB population;

(2) selection effects due to the GRBM trigger logic and the GRBM
+ WFC combined sensitivities and data truncation effects
 may introduce biases in our results;

(3) the observed correlations 
are a manifestation of the intrinsic properties of the 
population of the long GRBs, which result to be brighter and more energetic
(higher $E_p$) at larger distances.

In principle, we cannot reject any of these hypotheses. However, in the next
section we demonstrate that the selection effects, in case they are 
present, do not significantly influence our results.

\subsection{Selection and data truncation effects}

The evaluation of the biases due to selection effects and 
data truncation (e.g., the bias in the spectral parameters estimates due
to the detector finite bandwidth) is a topical issue in the study of the
correlations between different properties of GRBs,
as discussed by various authors (e.g.\cite{lloyd2000}). For instance,  
the combined sensitivities and energy bands of the WFC and GRBM could favor
the detection of harder and brighter GRBs at higher redshifts, thus
originating the $E_{rad}$ vs. (1+z) and the $E_p$ vs. $E_{rad}$ 
relations that we found in our sample.

\begin{figure}
\hspace{-0.73cm}\psfig{figure=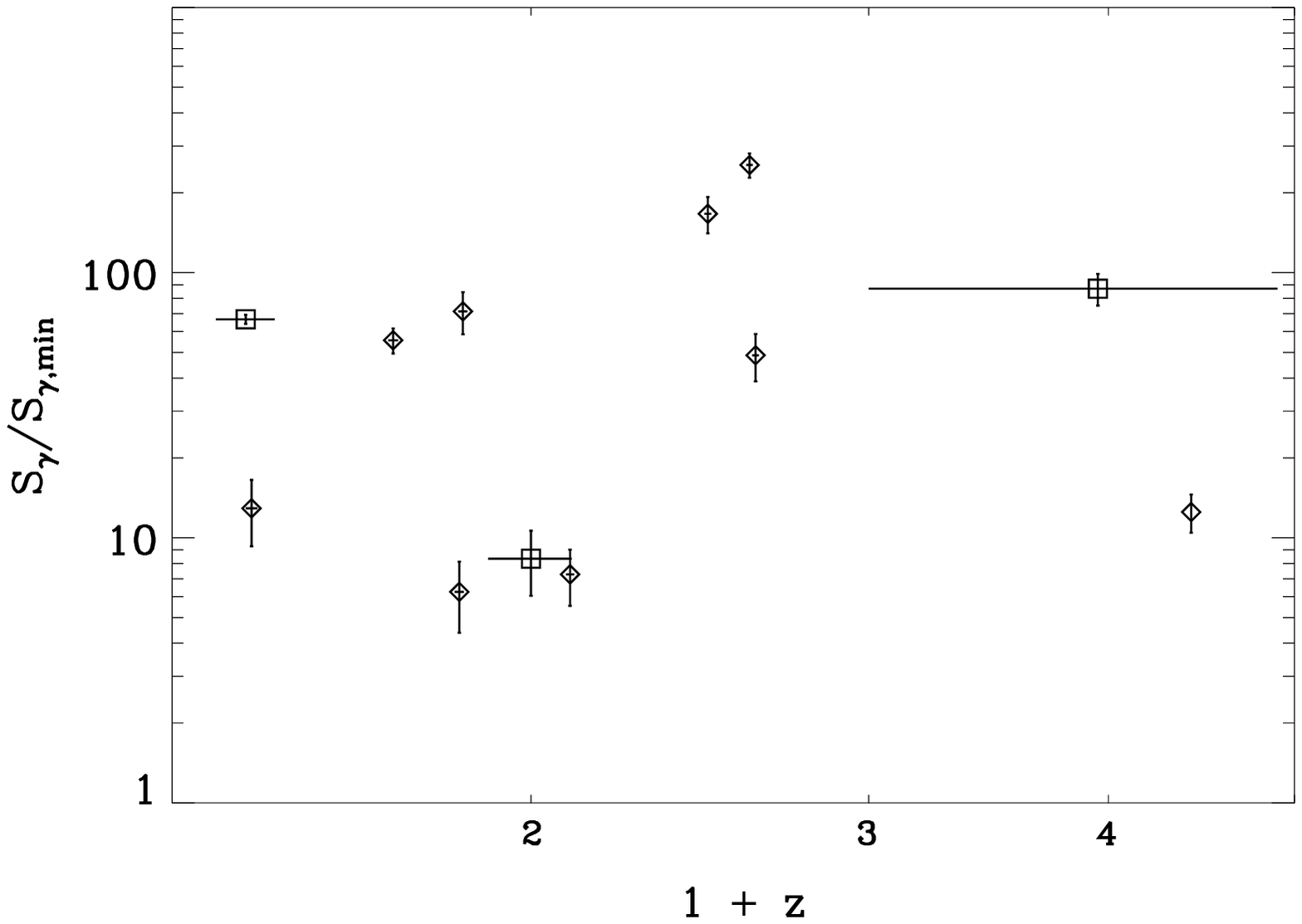,width=9.5cm,angle=0}
\caption{Ratio between the measured gamma--ray (40--700 keV) fluence and its minimum detectable
value for each GRB in our sample, plotted as a function of redshift.
Diamonds indicate the 9 GRBs with firm redshift
estimates, while squares refer to GRB~980326, GRB~980329 and GRB~000214.
}
\label{f:ratio}
\end{figure}

Feroci et al. (1999) performed an investigation on selection effects in a
sample of $\sim$15 GRBs detected with both GRBM and WFC with negative
results, apart the selection of long GRB ($\ge$6 s), as discussed above.
The fluences of most GRBs in our sample are well above their minimum
values as can be seen from  Table~\ref{t:tab1} and  Fig.~\ref{f:ratio}, where the 
ratio between the measured GRB fluence in 40--700 keV and its minimum detectable value $S_{\gamma,min}$
is shown as a function of $1+z$. Also, no correlation is visible
between the $S_{\gamma}/S_{\gamma,min}$ ratio and $1+z$. Even more distant GRBs have fluences much higher
than the minimum ones, also taking into account the GRB shape as discussed in Section 2.
Thus, we conclude that the observed trend between 
$E_{rad}$ and $1+z$ is likely not affected by selection effects.

Somewhat more complex is the evaluation of possible
data truncation biases in the correlations between spectral parameters
and intensity or redshift. Lloyd and Petrosian (1999) discuss extensively
this topic and account for data truncation in their correlation analysis 
between $E_p$ and bolometric fluence of a sample
of bright BATSE events. 
Following a method similar to the one used by them, we computed for each GRB
in our sample the minimum and maximum values of $E_p$ and $\alpha$, the
two spectral parameters for which we found significant correlation
with $E_{rad}$ and (1 + z), respectively. This is done by varying, first
downward and then upward, the spectral parameter of interest
while keeping fixed the other spectral parameters, the 
total radiated energy and the redshift, until the observer's rest--frame fluence reaches
the minimum detectable value in the WFC 2--28 keV or in the GRBM 40--700 keV energy band.
Data truncation will affect significantly
the correlation results if many values of the spectral parameters of
interest are very close to their minimum or maximum value
outside the acceptance range of other events. 
For the GRBs in our sample, none of the above conditions are verified,
and thus our results appear not significantly biased by data truncation. \\
Finally, we note that the events in our sample
are those for which an X--ray/optical follow--up has been performed.
The follow--up observations were not performed by selecting special GRBs
and thus we expect that no bias is introduced in the GRB sample selection.
%
%

\subsection{Consequences of the found correlations}

The possibility of a positive correlation between GRBs  peak energy $E_p$
and bolometric fluence was already inferred by several studies, mainly
based on BATSE detected events (e.g.  Mallozzi et al. 1995, Brainerd 1997),
which suffered the lack of the knowledge of their redshift 
(Lloyd et al. 2000). Our finding of an $E_p$ vs. $E_{rad}$
correlation confirms the previous results, with the advantage that the redshifts
of the GRBs in our sample are known. We find that $E_p$ depends on $E_{rad}$
accordingly to the following relationship:
\begin{equation}
       $$E_p \propto E_{rad}^{0.52\pm0.06}$$
\end{equation}

As discussed by Lloyd et al. (2000), a similar power law relation
($E_p \propto \epsilon_{rad}^{0.5}$)  is  expected in the case of an
Optically Thin Synchrotron Shock Model (OTSSM) for an electron distribution
with a power--law shape ($N(\gamma) = N_0 \gamma^{-p}$for
$\gamma > \gamma_m$) with $\gamma_m$, GRB duration and $N_0$ invariant from
burst to burst. We know that these assumptions are not completely true
(e.g. GRB duration changes from burst to burst) and $E_{rad}$ could be lower if
the emitted radiation is beamed, but this result should be taken into account
in working-out GRB emission models.

The interpretation of the other found relationship:
\begin{equation}
$$|\alpha| \propto (1+z)^{-0.78 \pm0.18}$$
\end{equation}
is not straightforward.

The fact that most of the low-energy photon indices $\alpha$
are in the range from $-2/3$ (instantaneous index predicted by the OTSSM;
e.g., Tavani 1996) and $-3/2$ (index predicted for synchrotron cooled leptons;
e.g.,  Cohen et al. 1997) is a strong hint in favor of the
the synchrotron as primary emission model. However equation (2) would suggest
that radiative cooling occur more actively in GRBs with relatively small
redshift. Apart the fact that this evolutionary effect is difficult to
justify, we have investigated other possible origin of the relation (2).
As a result, we have found that it does not reflect a
physical property of GRBs but is a consequence of other correlations.
In fact, also an evidence of correlation between $\alpha$ and $E_p$
was found (see Table~\ref{t:tab3}). This evidence was also found
for BATSE GRBs and was explained (e.g., Lloyd and Petrosian 2000) as due
partially to the dependence of the  $\alpha$ estimate on the spectral
curvature nearby $E_p$  and partially to the fact that 
$E_p = (2 + \alpha) \times  E_0$ (Band et al. 1993). Taking into account
the other found correlations between $E_p$ and $E_{rad}$ and between
$E_{rad}$ and $(1 + z)$, the expected relationship between $\alpha$ and
$(1+z)$ is given by $|\alpha| = (2.76\pm0.09) \times (1 + z)^{-0.75\pm0.06}$, 
with a power-law index which is fully consistent with the best fit
power--law index of equation (2).

\acknowledgements This research was partly supported by the
Italian Space Agency (ASI). We thank the teams of the BeppoSAX
Operative Control Center and Scientific Data Center for their
efficient and enthusiastic support to the GRB alert program.

\end{document}